\documentstyle[12pt]{article}
\begin{document}
\begin{center}
Comment on ``Statistical Mechanics of Non-Abelian Chern-Simons Particles"\\
\bigskip
by\\
\bigskip
C. R. Hagen\\
\medskip
Department of Physics and Astronomy\\
University of Rochester\\
Rochester, NY  14627
\end{center}

The second virial coefficient $B_2(T)$ for particles which interact with each
other through a Chern-Simons type coupling has been of considerable
interest in recent years.  In particular its form has been derived for
both the spin zero [1] and spin-1/2 cases [2] as a function of the flux
parameter $\alpha$.  More recently results have been obtained for the
case of a gas of spinless non-Abelian Chern-Simons particles.  [3]  It is the
purpose of this note to point out that the form of the virial
coefficient obtained in that work for the SU(2) case is quantitatively
incorrect and also that there exists a periodicity in the flux parameter
just as in the Abelian theory.  No such periodicity was noted in ref. 3.

To demonstrate the result one can begin with Eq. (25) of ref. 3 which reads
$$H^\prime_j = - {1\over 2\mu} \left[ {\partial^2\over \partial r^2} +
{1\over r} {\partial\over \partial r} + {1\over r^2}
\left({\partial\over\partial\theta} + i\omega_j\right)^2\right]$$
where $\omega_j$ can (by trivial scaling of the interaction parameter) be
taken to be
$$\omega_j = 2\alpha [j(j+1) - 2\ell (\ell +1)]\eqno(1)$$
where $\ell = 0, {1\over 2}, 1, ...$ and $j$ is any allowed angular
momentum (i.e., $j = 0, 1, 2,...2\ell)$. One infers the second virial
coefficient from the expression
$$ B_2(\alpha, T) = {1\over (2\ell +1)^2} \sum^{2\ell}_{j=0} (2j+1)
\left[{1+(-1)^{j+2\ell}\over 2} B^B_2(\omega_j,T) +
{1-(-1)^{j+2\ell}\over 2}B^F_2(\omega_j,T)\right]\eqno(2) $$
where superscripts $B$ and $F$ have been used to denote the virial
coefficients for the Abelian bosonic and fermionic cases, respectively.
It should be noted that the factors of $(-1)^{2\ell}$ have been
improperly omitted in ref. 3.  These factors are necessary to ensure
that the symmetric [antisymmetric] isospin states contribute only to
$B^B_2(\alpha,T)[B^F_2(\alpha,T)]$.

One now makes use of the well known results for spin zero particles
\[ B^{B(F)}_2(\alpha,T) =
{1\over 4}\lambda^2_T \left\{ \begin{array}{ll}
                  -1 + 4\delta - 2\delta^2 & \mbox{N even (odd)} \\
                  1 - 2\delta^2 & \mbox{N odd (even)}
                  \end{array}
                  \right. \]
where $\lambda_T$ is the thermal wavelength and $\alpha = N+\delta$ with
$N$ an integer such that $0 \leq \delta < 1$.  This leads upon insertion
into (2) the equation
$${4\over \lambda_T^2}B_2(\alpha, T) = \mp {1\over 2\ell +1} +
{2\over (2\ell +1)^2} \sum^{2\ell}_0 (2j+1)\delta_j
[1\pm(-1)^{j+2\ell}-\delta_j]\eqno(3)$$
where $\delta_j$ is given by
$$\omega_j = N_j + \delta_j$$
with $N_j$ an integer and $\omega_j$ given by Eq. (1).  The upper and lower
signs in (3) refer to the cases of even and odd $N_j$ respectively. [4]
It is important to note that the definition of $\omega_j$ implies for all
$j$ that $\omega_j$ is an integer multiple of $\alpha$.

One immediate point of contrast between (3) and the results of ref. 3 is
that for $\alpha \to 0$ the latter predicts that the virial coefficient
becomes the free bosonic result (see Eq. (30) of ref. 3).
This could only be correct if the
configuration space wave functions for all states were symmetric.  In
fact those states with angular momentum $j-1$, $j-3$, etc. necessarily
have antisymmetric wave functions and thus give a contribution to
$B_2(\alpha = 0, T)$ which is equal and opposite to that of the bosonic
case.  In fact a simple calculation for the free particle case easily
confirms the $\alpha = 0$ prediction of Eq. (3).

A final observation has to do with the periodicity issue.  Since, as
observed earlier, $\omega_j$ is an integer multiple of $\alpha$, it
follows that the contribution of each $j$ to $B_2(\alpha, T)$ is
periodic, with each recurrence occurring with (at most) a change of two
units in $\alpha$.  For specific $\ell$ values the period can, of course,
be seen to be reduced to a shorter interval.  Clearly the total virial
coefficient also satisfies this periodicity condition, thereby showing
that there is little qualitative difference between the Abelian and
non-Abelian cases.
\bigskip

\noindent Acknowledgement

This work has been supported in part by a grant from the  U. S. Dept. of
Energy DE-FG02-91ER40685.
\bigskip

\noindent References

\begin{enumerate}
\item D. P. Arovas, R. Schrieffer, F. Wilczek, and A. Zee, Nucl. Phys. B
{\bf 251}, 117 (1985); A. Comtet, Y. Georgelin, and S. Ouvry, J. Phys.
A{\bf 22}, 3917 (1989).
\item T. Blum, C. R. Hagen, and S. Ramaswamy, Phys. Rev. Lett. {\bf 64},
709 (1990).
\item T. Lee, Phys. Rev. Lett. {\bf 74}, 4967 (1995).
\item It is worth noting that in ref. 3 no differentiation is made between
even and odd $N_j$.  This is not permissible even in the Abelian case.

\end{enumerate}

\end{document}